\documentclass[aps,prd,amsmath,amssymb,amsfonts,nofootinbib,superscriptaddress,twocolumn]{revtex4-1}
\usepackage[utf8]{inputenc}
\usepackage{graphicx}% Include figure files
\usepackage{subfigure}
\usepackage{datatool}
\usepackage[scientific-notation=true]{siunitx}
\usepackage[dvipsnames]{xcolor}
\usepackage{xspace}
\usepackage{orcidlink}
\usepackage{url}
\def\UrlBreaks{\do\/\do-}
\usepackage{breakurl}
\usepackage{longtable}
\usepackage[shortlabels]{enumitem}
\usepackage[capitalise]{cleveref}
\def\pul{pulsar\xspace}
\def\puls{pulsars\xspace}

\def\gwh{GW\xspace}
\def\cwh{continuous-wave\xspace}
\def\cws{continuous waves\xspace}
\def\gws{GWs\xspace}

\def\ns{neutron star\xspace}
\def\nss{neutron stars\xspace}
\def\nsh{neutron-star\xspace}

\def\Nss{Neutron stars\xspace}
\def\et{Einstein Telescope\xspace}

\def\fh{frequency-Hough\xspace}
\def\lvk{LIGO, Virgo and KAGRA\xspace}
\def\D{DECIGO\xspace}

\def\BD{Brans-Dicke\xspace}
\def\sgwb{stochastic gravitational-wave background \xspace}
\def\nn{\nonumber}

\def\snr{signal-to-noise ratio\xspace}
\def\eos{equation-of-state\xspace}
\def\EM{electromagnetic\xspace}
\def\psd{power spectral density\xspace}
\def\psds{power spectral densities\xspace}
\def\l{\left(}
\def\r{\right)}
\def\co{compact object\xspace}
\def\coss{compact objects\xspace}
\def\eco{exotic compact object\xspace}
\def\ecos{exotic compact objects\xspace}
\newcommand{\bea}{\begin{eqnarray}}
\newcommand{\eea}{\end{eqnarray}}
\newcommand{\be}{\begin{equation}}
\newcommand{\ee}{\end{equation}}

\newcommand{\TFFT}{T_\text{FFT}}

\newcommand{\Tobs}{T_\text{obs}}

\newcommand{\fgw}{f_\text{gw}}
\newcommand{\fdotmax}{\dot{f}_{\rm max}}
\newcommand{\fdotgw}{\dot{f}_{\rm gw}}
\newcommand{\Izz}{I_\text{zz}}
\newcommand{\hmin}{h_\text{0,min}}

\newcommand{\tcr}{}
\newcommand{\tcb}{}

\newcommand{\unitfdot}{\text{ Hz/s}}
\newcommand{\unitf}{\text{ Hz}}
\newcommand{\unitdist}{\text{ kpc}}
\newcommand{\unitizz}{\text{ kg $\cdot$ m$^2$}}
\newcommand{\unitD}{\text{ kg $\cdot$ m}}

\graphicspath{{./figures/}}

\begin{document}

\title{Searching for continuous gravitational waves from highly deformed compact objects with DECIGO}

\author{Andrew L. Miller\,\orcidlink{0000-0002-4890-7627}}
\email{andrew.miller@nikhef.nl}
\affiliation{Nikhef -- National Institute for Subatomic Physics,
Science Park 105, 1098 XG Amsterdam, The Netherlands}
\affiliation{Institute for Gravitational and Subatomic Physics (GRASP),
Utrecht University, Princetonplein 1, 3584 CC Utrecht, The Netherlands}
\author{Federico De Lillo\,
\orcidlink{0000-0003-4977-0789}}
\email{federico.delillo@uantwerpen.be}
\affiliation{Universiteit Antwerpen, Prinsstraat 13, 2000 Antwerpen, Belgium}

\date{\today}

\begin{abstract}

Searches for continuous gravitational waves from isolated compact objects and those in binary systems aim to detect non-axisymmetric, deformed neutron stars at particular locations in the Galaxy or all-sky. However, a large fraction of known pulsars have rotational frequencies that lie outside the audio frequency band, rendering current detectors insensitive to these pulsars. In this work, we show that DECIGO, a future space-based deci-hertz gravitational-wave interferometer, will be sensitive to severely deformed compact objects, e.g. hybrid stars, neutron stars, or magnetars. We estimate the number of possible compact objects that could be detected with such high deformations, both via their individual continuous gravitational-wave emission and the stochastic gravitational-wave background created by a superposition of gravitational waves from the $\sim 10^8$ compact objects in the Galaxy. 
Furthermore, we show that the existence of such compact objects could be probed across a wide parameter space at a fraction of the computational cost of current searches for isolated compact objects and those in binary systems. For known pulsars, we will be able to both beat the spin-down limit and probe the Brans-Dicke modified theory of gravity parameter $\zeta<1$ for approximately 85\% of known pulsars with  $\fgw<10$ Hz, the latter of which is currently only possible for $\mathcal{O}(10)$ pulsars. DECIGO will thus open a new window to probe highly deformed compact objects.

\end{abstract}

\maketitle

\section{Introduction}

Continuous gravitational waves are expected to arise from deformed, asymmetrically rotating \nss, and last for durations that significantly exceed the observation times of gravitational-wave (\gwh) detectors such as \lvk \citep{2015CQGra..32g4001L,2015CQGra..32b4001A,Aso:2013eba}. While not yet detected, such \gws could contain interesting information regarding the \eos, and also answer questions regarding the nature of matter within \ns.

Extensive efforts have targeted both known and unknown \nss, resulting in competitive constraints on the degree of deformation on \nss, known as the ``ellipticity'' \citep{Sieniawska:2019hmd,Tenorio:2021wmz,Riles:2022wwz,Piccinni:2022vsd,Miller:2023qyw,Wette:2023dom}. While \emph{targeted} searches for \cws from known \puls could detect ``mountains'' as tiny as tens of micrometers \citep{LIGOScientific:2020gml,LIGOScientific:2020lkw,LIGOScientific:2021hvc,Ashok:2021fnj,Ashok:2024fts}, they are limited in scope to \coss whose existence we can already infer through \EM radiation. On the flip-side, we expect $10^{8}$ electromagnetically dark \nss to exist in the galaxy \citep{sartore2010galactic}. Thus, all-sky searches that attempt to find \nss anywhere in the sky are performed \citep{KAGRA:2022dwb,Steltner:2023cfk}, but suffer sensitivity losses due to an extremely wide parameter space -- unknown frequency, spin-down and sky position -- that makes deep searches extremely computationally intensive. 

Coupled with the strengths and limitations of targeted and all-sky \cwh searches are astrophysical uncertainties: the a priori unknown spin-frequency, ellipticity, and spatial distributions of \nss in the galaxy \citep{Reed:2021scb}. While the Australia Telescope National Facility (ATNF) catalog provides some evidence of a bi-modal spin frequency distribution around a few Hz and a few hundreds of Hz \citep{Manchester:2004bp}, it is by far not an exhaustive catalog of what \nss could actually exist, but is the most observationally motivated choice we can make for these distributions. 

Results from targeted and all-sky searches for \nss constrain ellipticities anywhere between [$10^{-7}$, 1], depending on the distance to the source, the frequency, and the inclination angle assumed \citep{Wette:2023dom}. In all-sky searches, lower frequencies in particular suffer from weaker constraints than those at higher frequencies \citep{KAGRA:2022dwb,Steltner:2023cfk}, due to the scaling of the \gwh amplitude with the square of the frequency and decreased detector sensitivity below, say, 50 Hz compared to 100 Hz \citep{maggiore2008gravitational}. Moreover, theoretical predictions for the existence of such largely deformed \nss, e.g. magnetars, \citep{DallOsso:2008kll,Cheng:2015rja,Mastrano:2011tf,Suvorov:2023mqt,Suvorov:2023nxi}, and  \ecos, e.g. \nss with quark cores (hybrid stars) \citep{Owen:2005fn, Haskell:2007sh, Glampedakis:2012qp}, solid strange quark stars \citep{Owen:2005fn} and Thorne-Zytkow objects (TZOs) \citep{thorne1975red,thorne1977stars,DeMarchi:2021vwr}, allow for the possibility of ellipticities around $[10^{-4},1]$ depending on the assumed \nsh mass, radius, and maximum crustal strain \tcr{in Eq. 7 of \cite{Owen:2005fn} (up to 0.1 \citep{Horowitz:2009ya}). In particular, the known magnetar 4U 0142+61 could have an ellipticity of $\sim 1.6\times 10^{-4}$ \cite{Makishima:2014dua}, highlighting that even for standard scenarios, large deformations remain possible.}

\tcr{There have been many years debate in the literature about the maximum ellipticity that \nss could sustain. Early studies predicted maximum ellipticities of $\mathcal{O}(10^{-6})$ \cite{Ushomirsky:2000ax,Haskell:2006sv,Johnson-McDaniel:2012wbj}, though the authors in \cite{Gittins:2020cvx} have cast doubt on previous calculations of maximum deformation size and find an order of magnitude smaller mountain size. However, the work in \cite{Gittins:2020cvx} only concerns canonical \nss and make assumptions regarding the form of the force that deforms the crust, the model of crust as an elastic solid, and the requirement that the breaking strain of the crust is reached only at one point. A way to connect the studies that predict larger mountains \cite{Ushomirsky:2000ax,Haskell:2006sv,Johnson-McDaniel:2012wbj} and that in \cite{Gittins:2020cvx} would be to perform simulations of the evolution of \nss across cosmic time to see how their mountains are built, as argued in \cite{Gittins:2020cvx}. Likewise, others \cite{Morales:2022wxs} argue that \nss could sustain mountains larger than those predicted in \cite{Gittins:2020cvx}, consistent with previous work, and offer a comparison of the maximum ellipticity that fluid stars, solid-fluid stars and the crust could handle, which spans $\mathcal{O}(10^{-8})-\mathcal{O}(10^{-2})$. Thus, the possibility of highly-deformed \nss, and of other kinds of stars, such as strange stars, has not yet been definitively ruled out: in particular, strange stars could have ellipticities as high as $10^{-1}$ \cite{Owen:2005fn}.}

\tcr{The existence of a variety of theoretical models for canonical \nss, strange stars and \ecos that permit large mountains, the lack of experimental constraints on high-ellipticity sources, and the presence of a peak around 1 Hz in the frequency distribution of known \puls \cite{Manchester:2004bp}, motivate the need to consider the possibility of such sources rotating at low frequencies that \D or \et could detect.}

So far, \nss with spin frequencies of at least 5 Hz have been searched for using \lvk data \cite{LIGOScientific:2020qhb,KAGRA:2022dwb,Steltner:2023cfk,Covas:2024nzs}; here, however, we point out the possibility to search for known \puls and unknown \nss whose spin frequencies are below 5 Hz using future space-based \D data. \D \citep{Kawamura:2020pcg} is planned to be a triangular Japanese space-based laser interferometer detector with 1000 km-long arms in orbit the Earth that is exquisitely sensitive to the deci-hertz \gwh frequency range (0.1-10 Hz). If the ATNF catalog indicates in some way the true spin frequency distribution of \nss, then \D will be able to probe \nsh frequencies inaccessible to current ground-based detectors, which comprise most of the known \nss.

\tcr{As we will show, \D will allow us to be sensitive to \coss whose deformations are an order of magnitude larger than those of this magnetar, which has also been studied concurrently in \cite{10.1093/mnras/staf774,10.1093/mnras/staf704}. }
If such highly-deformed \coss did exist, a recent analysis concludes that less than tens of thousands of these objects could be within our Galaxy if they have $\varepsilon>10^{-6}$ \cite{Prabhu:2024lzl}. However, this work assumes only \gwh frequencies above $\sim20$ Hz and implicitly works in the small ellipticity limit, since it employs results from a recent search \citep{KAGRA:2022dwb} that limits $|\dot{f}|<10^{-8}$ Hz/s at frequencies up to 2000 Hz, which thus restricts the maximum probable ellipticity (see \cref{eqn:fdotgw}). Extremely deformed \nss, especially at high frequencies, thus actually lie \emph{outside} of the parameter space searched over in \cite{KAGRA:2022dwb}. \tcb{Despite limited evidence for their existence, such highly deformed \coss remain valuable targets for \D, given the potential to discover unexpected astrophysical systems.}

% The theoretical predictions for highly deformed \nss, the presence of a peak around 1 Hz in the frequency distribution of known \puls, and the limitations of current searches to actually detect such ``\ecos'', motivate the need to consider alternative ways of probing such systems. The proposed space-based GW interferometer \D \citep{Kawamura:2020pcg}, which bridges the gap between the mHz to the few-Hz band, provides a way of probing such systems.

Even though searches for isolated \nss have historically been the focus in the \cwh community, a vibrant effort also exists to detect \cws from \nss in binary systems, whose signals would be modulated by the orbital motion of their companions. Such searches are typically motivated by the fact that around half of \puls that we observe exist in binary systems, and by the existence of low-mass X-ray binaries such as Scorpius X-1 \citep{Giacconi:1962zz} that are expected to be strong emitters of \gws \citep{Watts:2008qw}. However, these searches suffer greatly from an enlarged parameter space --semi-major axis, orbital period, time of ascension, and eccentricity should in principle be searched over \citep{Leaci:2015bka,Covas:2019jqa} -- that permits only limited efforts to probe the orbital parameters of the binary \citep{LIGOScientific:2020qhb,Covas:2024nzs}. To address this problem, \D will provide a way to easily probe the existence of \nss in binary systems at a fraction of the computational cost used now, while providing even greater coverage of the orbital parameter space, due to the fact that the computational cost of doing these searches for unknown \nss, isolated or in binary systems, scales immensely with the \gwh frequency \citep{Prix:2009oha}.

We thus divide this paper into different ways to search for \nss using future \D data and the projected constraints from those searches. To begin, we provide a brief introduction to the \gwh signal from isolated \nss and those in binary systems in Sec. \ref{sec:gwinsp}. In Sec. \ref{sec:cw-lims}, we discuss the possible constraints on ellipticity and the \BD theory of gravity that could be obtained from both targeted and all-sky searches for \nss. Then, in Sec. \ref{sec:allsky}, we explain how to calculate the number of deformed \nss that all-sky searches could be sensitive to in the \D era as a function of ellipticity. We then show in Sec. \ref{sec:nsbin} that \cwh searches could, with current allocations of computing power, explore a much wider parameter space of \nss in binary systems while also being able to handle highly eccentric systems without the need to explicitly search over eccentricities. In Sec. \ref{sec:sgwb}, we determine to what extent we could be sensitive to the \sgwb arising from \cws being emitted by isolated \nss at \D frequencies, in case such systems could not be detected individually. We conclude in Sec. \ref{sec:concl} with ideas for future work and thoughts about the prospects of \cwh searches with \D.

\section{Gravitational waves from deformed compact objects}\label{sec:gwinsp}

In the canonical model, a deformed, non-axisymmetric, rotating \ns (or any \eco)  will have a time-varying quadrupole moment, and thus emit \gws as it rotates. The gravity on the surface of the \co is huge with respect to that on Earth, and thus distortions are conventionally assumed to be small, of $\mathcal{O}($mm) or less. Assuming \gwh dominates the spin-down of the \co with respect to electromagnetic radiation, rotational power is converted completely to \gwh power, spinning down the \ns at a rate $\fdotgw$ given by \cite{maggiore2008gravitational}:

\begin{align}
\fdotgw &= -\frac{32 \pi^4 G}{5 c^5} \Izz \varepsilon^2 \fgw^5 \nn
\\
&=-1.71\times 10^{-18} \unitfdot \l\frac{\Izz}{10^{38} \unitizz }\r \nn \\ &\times  \l\frac{\varepsilon}{10^{-3}}\r^2 \l\frac{\fgw}{1\unitf}\r^5
\label{eqn:fdotgw}
\end{align}
where $\fgw$ is the \gwh frequency (twice the rotational frequency), $\Izz$ is the principal moment of inertia along the $z$-axis of the \ns, $\varepsilon\equiv \frac{I_{\rm xx}-I_{\rm yy}}{\Izz}$ is the ellipticity, $G$ is Newton's gravitational constant, and $c$ is the speed of light.

The duration of this \gwh signal for the parameters given in \cref{eqn:fdotgw} is roughly $\tau_{\rm gw} \sim f/\fdotgw\sim 10^{10}$ years, much longer than the observing run of any \gwh detector; thus, these signals are quasi-monochromatic and persistent in time. The \gwh signal can therefore be approximated as Taylor series expansion of the signal frequency evolution to linear order, neglecting higher-order derivatives in frequency due to their minute effects on $\fgw(t)$:

\begin{align}
    \fgw(t) &= \l f_{0} +\fdotgw(t-t_0)\r \nn \\ 
    &\times \left(1 + \frac{\vec{v}(t) \cdot \hat{n}}{c} 
    - a_{p} \Omega \cos\left[\Omega (t-t_{\textrm{asc}})\right] \right)\;.
    \label{eq:track}
\end{align}
\cref{eq:track} also encodes frequency modulations due to the motion of the detector with respect to the source, and the orbital motion of the \ns around its companion if it exists in a binary.
$\vec{v}(t)$ is the detector velocity with respect to the solar-system barycenter, $\hat{n}$ is the vector connecting the detector to the source, $a_p$ is the semi-major axis expressed in units of light seconds (l-s), $\Omega$ is the orbital angular frequency, and $t_{\rm asc}$ is the time of ascension. The amplitude of the signal can be written as \citep{maggiore2008gravitational}:

\begin{align}
    h_0 &= \frac{4 \pi^{2} G}{c^4} \frac{\varepsilon \Izz \fgw^2}{d} \nn \\
    &= 1.05\times 10^{-27} \l\frac{1 \unitdist}{d}\r \l \frac{\varepsilon}{10^{-3}}\r \nn \\ &\times \l \frac{\Izz}{10^{38}\unitizz}\r \l \frac{\fgw}{1\unitf}\r^2
    \label{eqn:h0}
\end{align}
where $d$ is the distance from the source.

A relevant quantity for targeted searches is the ``spin-down limit'', which quantifies the maximum \gwh amplitude assuming that all rotational power is converted to \gwh power:

\begin{align}\label{eq:h0sd}
    h_{0}^{\rm sd} &=\frac{1}{d}\left(\frac{5G\Izz}{2c^3}\frac{|\fdotgw|}{\fgw}\right)^{1/2} \nn \\
    &= 2.63\times 10^{-28} \l\frac{1 \unitdist}{d}\r \l\frac{\Izz}{10^{38}\unitizz}\r^{1/2} \nn \\ 
    &\times \l\frac{|\fdotgw|}{1.71\times 10^{-18}  \unitfdot}\r^{1/2} \l \frac{1\unitf}{\fgw}\r^{1/2}
\end{align}
Typically, upper limits from targeted searches are considered physically plausible if the amplitude that can be reached is lower than $h_{0}^{\rm sd}$.

We can also constrain modified theories of gravity, in particular \BD theory \citep{Brans:1961sx}, using \cws, as done in \citep{LIGOScientific:2021hvc}. This theory predicts a scalar polarization in addition to the two ordinary tensor ones in general relativity, and the dominant contribution to \gwh emission occurs from a time-varying dipole moment  \citep{Verma:2021nbz} at the rotational frequency of the star
Assuming that the dipole moment $D\sim MR$ in the reference frame of a \ns with mass $M$ and radius $R$ has only an $x$-component, the signal amplitude $h_0^d$ is 
\begin{align}\label{eqn:h0d}
    h_{0}^d &= \frac{4 \pi G}{c^3} \zeta \frac{D \fgw}{d} \nn \\
    &= 1.01\times 10^{-28} \l\frac{\zeta}{10^{-3}}\r \l \frac{D}{10^{29} \unitD}\r \nn \\ &\times\l\frac{1\unitdist}{d}\r \l \frac{\fgw}{1\unitf}\r
\end{align}
where $\zeta$ is the \BD parameter that quantifies the fraction of \gwh power that goes into the scalar mode.

\section{Projected sensitivities from continuous-wave searches}\label{sec:cw-lims}

\subsection{Limitations of current continuous-wave searches}

\begin{figure*}[ht!]
     \begin{center}
        \subfigure[ ]{%
            \label{fig:fdot_vs_f-ellip}

        \includegraphics[width=0.5\textwidth]{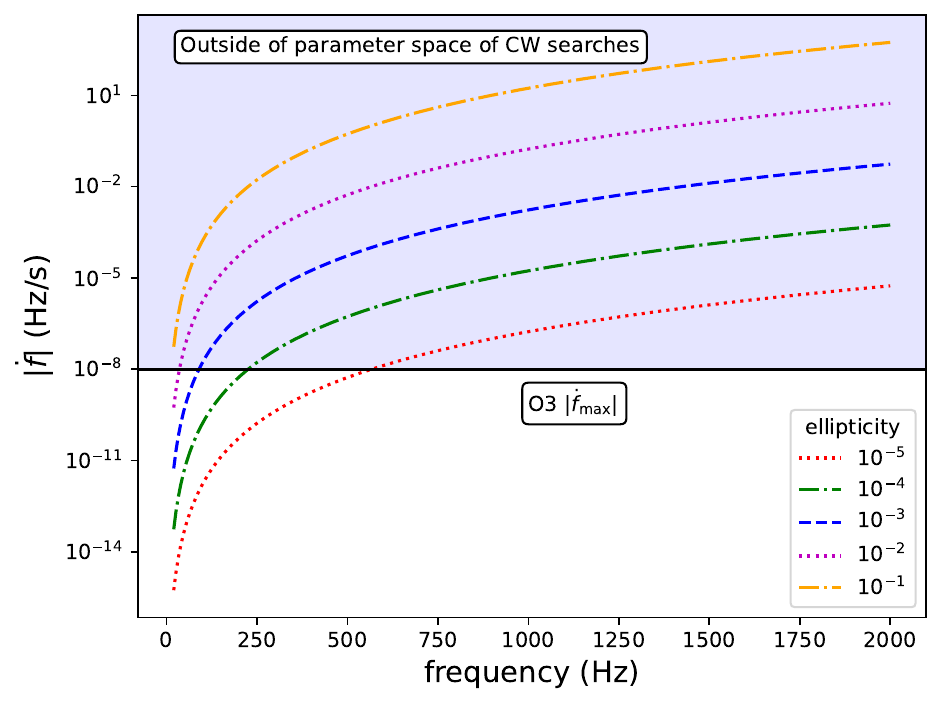}
        }%
        \subfigure[]{%
           \label{fig:max_ellip_vs_f}
           \includegraphics[width=0.5\textwidth]{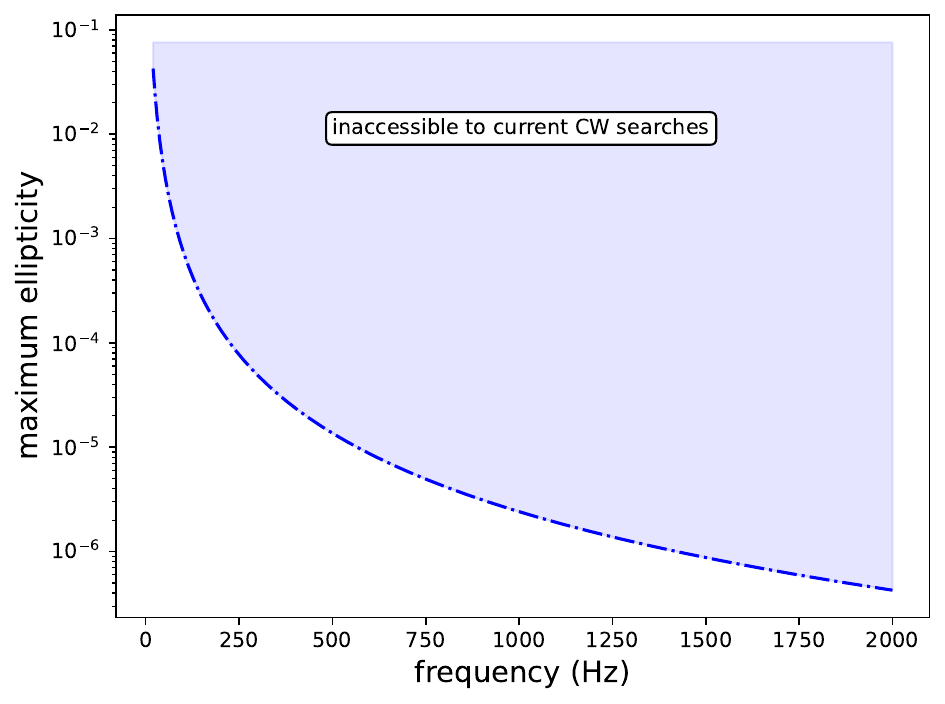}
        }\\ %  ------- End of the first row ----------------------%
    \end{center}
    \caption{Ellipticity range covered by current \cwh searches. (a) The frequency/spin-down parameter space for \coss with different ellipticities is shown here, along with a black line indicating the maximum spin-down to which current \cwh searches can be sensitive. The light blue region indicates spin-downs, and subsequently ellipticities, that current searches cannot detect. (b) The maximum ellipticity that current \cwh searches can be sensitive to assuming $|\fdotmax|=10^{-8}$ Hz/s. The blue region indicates directly which ellipticities are too high to be probed currently. In both plots, $\Izz=10^{38}$ \unitizz. } 
      \label{fig:inacc-ellip}
\end{figure*}

Current all-sky \cwh searches have been designed to be sensitive to small deformations on \nss, since canonical models of \nss do not predict large deviations from spherical symmetry \cite{Morales:2022wxs}. Such sensitivity to tiny ``mountains'' inherently implies a limitation on the maximum mountain size to which such searches are sensitive because these searches restrict the spin-down range that is analyzed, which thus limits the ellipticity that could be probed (see \cref{eqn:fdotgw}). In \cref{fig:fdot_vs_f-ellip}, we show that current \cwh searches can only detect ellipticities below $\sim 10^{-5}$ at almost all frequencies, highlighting that highly-deformed \nss may be more optimally searched for at deci-hertz frequencies. The black line in \cref{fig:fdot_vs_f-ellip} shows $|\fdotmax|=10^{-8}$ Hz/s \citep{KAGRA:2022dwb} to which current \cwh searches can be sensitive, which breaks the plot into two halves: the gray region is inaccessible to current \cwh searches, while the white region can be probed. Ellipticities larger than $\sim 10^{-3}$ cannot be probed above 100 Hz in current analyses. Viewed in another way, \cref{fig:max_ellip_vs_f} shows the maximum ellipticity that can be probed with current \cwh searches assuming $|\fdotmax|=10^{-8}$ Hz/s. \tcr{Additionally, Viterbi-based algorithms \cite{Viterbi:1967} that look for spin-wandering via Hidden Markov Models (HMMs) \cite{Bayley:2019bcb,Suvorova:2016rdc,Suvorova:2017dpm,Sun:2017zge} are in principle sensitive to random fluctuations of the \nsh spin frequency, but practically are restricted to a maximum $\dot{f}$ by the choice of coherence time, which still limits the maximum ellipticity to which they are sensitive \cite{KAGRA:2022dwb}.}

For completeness, we note that current \cwh searches could in principle look for higher ellipticity sources, but would need a strong astrophysical justification to warrant the increasing computing power required. Furthermore, these searches would likely have to go beyond one or two terms in the Taylor series expansion of the frequency (\cref{eq:track}), which would necessitate algorithmic improvements or potentially other methods to handle the full frequency evolution given by the integral of \cref{eqn:fdotgw}, e.g. \cite{Miller:2018rbg,Sun:2018owi,Oliver:2018dpt,Banagiri:2019obu}.  \tcb{We also highlight that in \cref{fig:inacc-ellip}, the minimum \gwh frequency currently analyzed is 20 Hz. However, these searches can be easily adapted to search for sources at \D frequencies because the spin-downs of even high ellipticity sources are much smaller at frequencies below 20 Hz compared to those at frequencies above 20 Hz. As we will argue, \cwh techniques can thus probe highly deformed compact objects in \D.}

\subsection{Targeted searches}

Targeted search methods can analyze the data coherently via the matched filter to obtain the maximum possible sensitivity to a signal buried in noise. Many different techniques have been developed over the years \citep{Dupuis:2005xv,Astone:2012uz,Jaranowski:1998qm,Jaranowski:2010rn}, which give very similar sensitivities to quasi-monochromatic signals.

To determine the sensitivity of matched filtering to \cws in \D, we employ the following estimate of the minimum detectable amplitude at the 95\% confidence level $\hmin^{95\%}$ from a Bayesian search \citep{Dupuis:2005xv}:

\begin{align}\label{eqn:h0min-bayes}
    \hmin^{95\%} &\simeq 10.8\sqrt{\frac{S_n(f)} { \Tobs}} \nn; \\
    &\simeq 1.57\times 10^{-27} \left(\frac{1\text{ year}}{\Tobs}\right)^{1/2}
\left(\frac{S_n(\fgw=1\text{ Hz})}{6.70\times 10^{-49} \text{ Hz}^{-1}}\right)^{1/2}
\end{align}
where $S_n(f)$ is the noise power spectral density and $\Tobs$ is the observation time.

\begin{figure*}[ht!]
     \begin{center}
        \subfigure[ ]{%
            \label{fig:ellip_proj_targ}

        \includegraphics[width=0.5\textwidth]{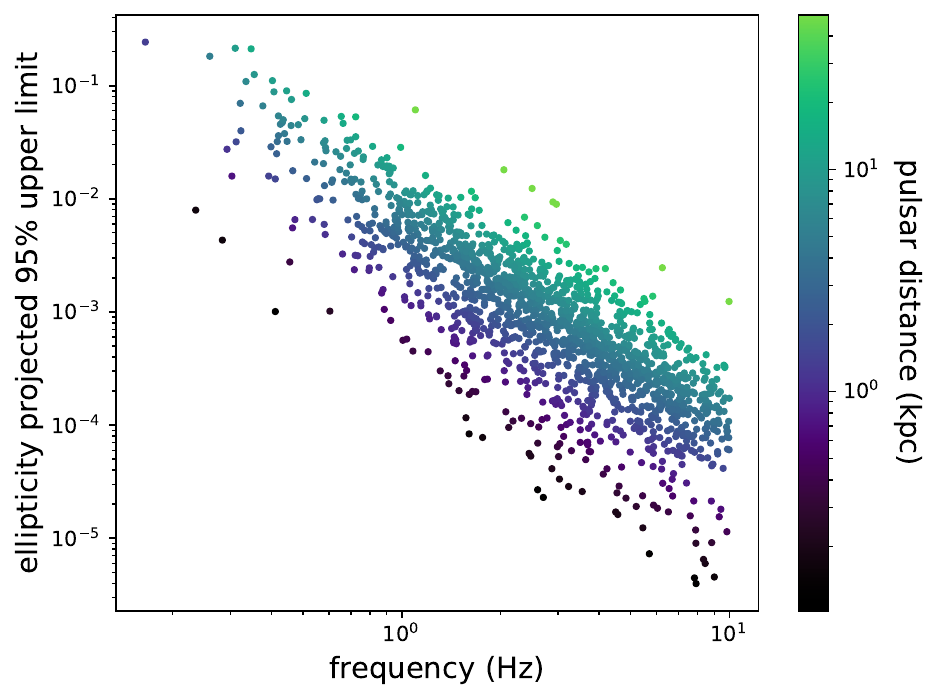}
        }%
        \subfigure[]{%
           \label{fig:bd_proj_targ}
           \includegraphics[width=0.5\textwidth]{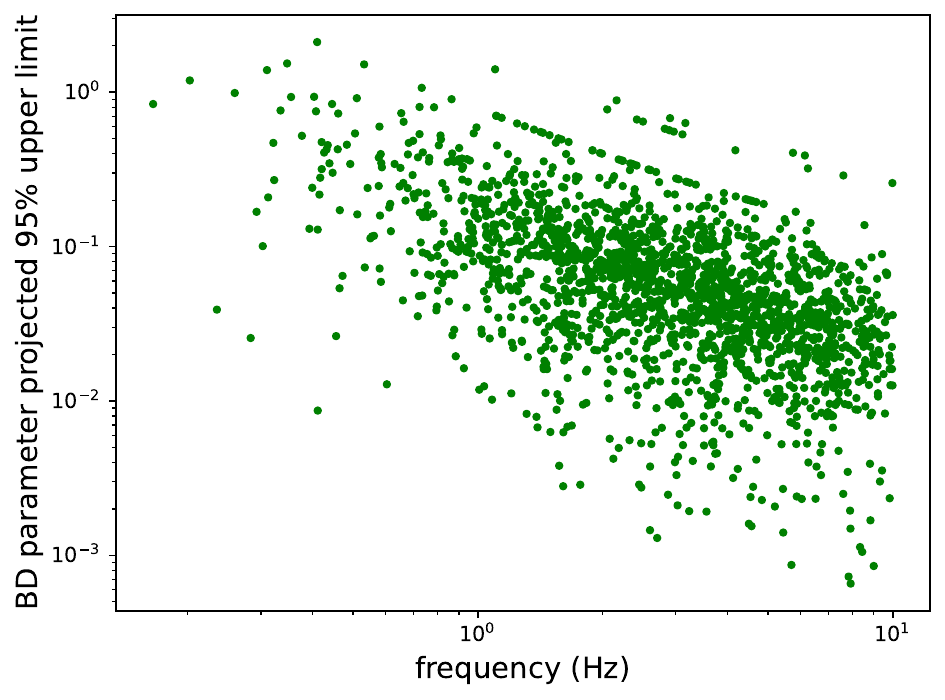}
        }\\ %  ------- End of the first row ----------------------%
    \end{center}
    \caption{Projected constraints from future targeted searches in \D on the (a) ellipticity upper limit and on the (b) parameter of \BD theory that encapsulates the fraction of \gwh energy going into dipole radiation.} 
     \label{fig:cw-targ}
\end{figure*}

Combining \cref{eqn:h0min-bayes,eq:h0sd,eqn:fdotgw}, we can obtain a projected upper limit on the detectable ellipticity at 95\% confidence. For the known \puls in the ATNF catalog, we plot in \cref{fig:ellip_proj_targ} these projected constraints as a function of \gwh frequency using the \D \psd \cite{Kawamura:2020pcg}. We only show here \puls whose spin-down limit can be surpassed in \D, which comprises about 85\% of \puls with $\fgw<20$ Hz. Higher frequencies permit more stringent constraints on the ellipticity because smaller deformations can be sustained more easily on more rapidly rotating \nss compared to slower rotating ones.

By combining \cref{eqn:h0min-bayes,eqn:h0d}, we can also project constraints on the \BD parameter $\zeta$ from targeted searches, which are shown in \cref{fig:bd_proj_targ}. Again, we have restricted this plot only to \puls for which we can constrain $\zeta<1$, which comprises about 97\% of known \puls whose rotation frequencies $f_{\rm rot}=\fgw<20$ Hz.

\subsection{All-sky searches}

All-sky \cwh searches in \D could be sensitive to slowly rotating \nss anywhere in the galaxy, although with a slightly reduced sensitivity with respect to targeted ones. Due to excessive computational costs, these semi-coherent searches break the data into chunks of length $\TFFT$ that are analyzed coherently, and then combined incoherently \cite{jaranowski1998data,Krishnan:2004sv,Astone:2014esa,Suvorova:2016rdc}. One particular method, the \fh \citep{Astone:2014esa}, would have a minimum detectable amplitude at 95\% confidence of:

\begin{align}
\hmin^{95\%}&\simeq
\Lambda\sqrt{\frac{S_n(\fgw)} {\TFFT^{1/2}\Tobs^{1/2}}} \nn \\
&\simeq 8.15\times 10^{-27} \left(\frac{1\text{ day}}{\TFFT}\right)^{1/4} \left(\frac{1\text{ year}}{\Tobs}\right)^{1/4} \nn \\ &\times
\left(\frac{S_n(\fgw=1\text{ Hz})}{6.70\times 10^{-49} \text{ Hz}^{-1}}\right)^{1/2} \left(\frac{\Lambda}{12.81}\right)
\label{eqn:h0min-fh}
\end{align}
where $\Lambda$ is a method-dependent parameter (see Eq. 67 in \citep{Astone:2014esa}). 

\begin{figure*}[ht!]
     \begin{center}
        \subfigure[ ]{%
            \label{fig:ellip_proj_all-sky}

        \includegraphics[width=0.5\textwidth]{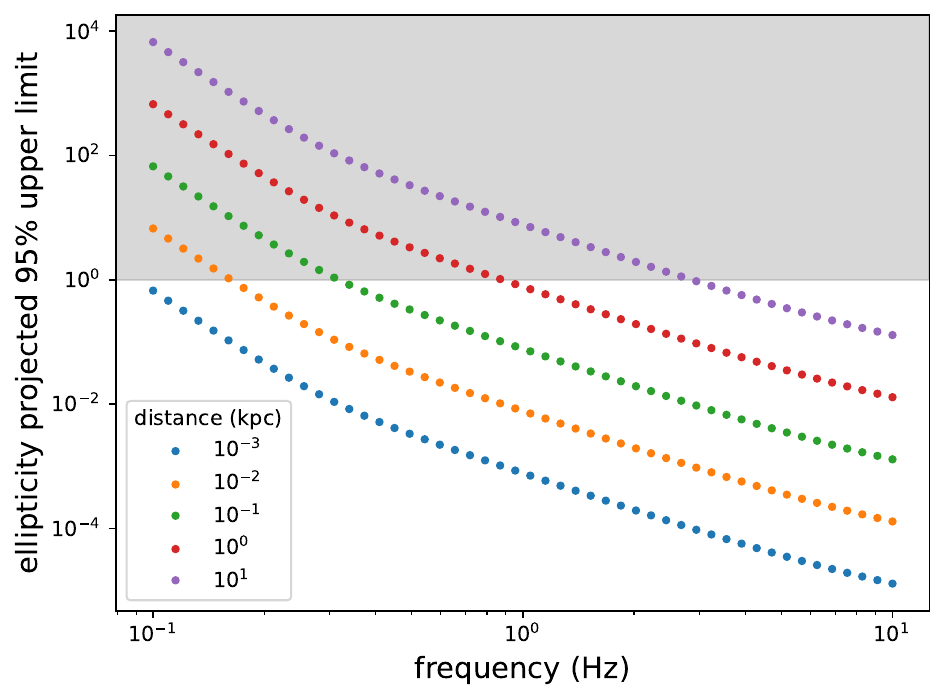}
        }%
        \subfigure[]{%
           \label{fig:bd_proj_all-sky}
           \includegraphics[width=0.5\textwidth]{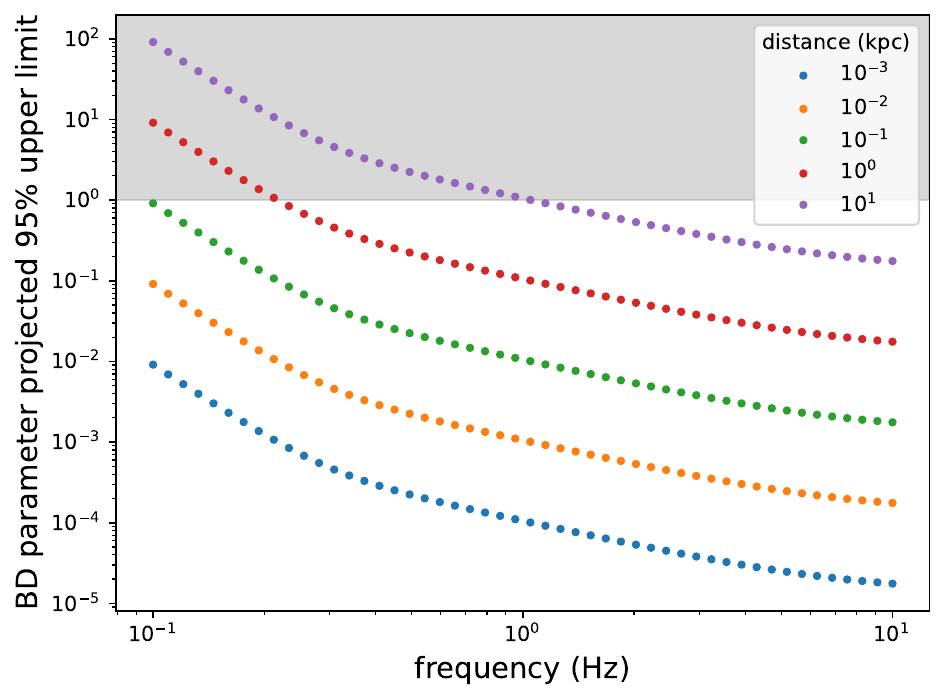}
        }\\ %  ------- End of the first row ----------------------%
    \end{center}
    \caption{Projected 95\% confidence-level constraints from future all-sky searches in \D on the (a) ellipicity upper limit and on the (b) parameter of \BD theory that encapsulates the fraction of \gwh energy going into dipole radiation. $\TFFT=1$ day, $\Tobs=1$ year, $\Izz=10^{38}$ \unitizz, $D=10^{29}$ \unitD. The gray shaded regions denote nonphysical ellipticities and \BD parameters.} 
     \label{fig:cw-all-sky}
\end{figure*}

Using \cref{eqn:h0min-fh,eqn:h0} and \cref{eqn:h0min-fh,eqn:h0d}, we can project constraints on ellipticity and $\zeta$ upper limits as a function of \gwh frequency and distance reach of the search, respectively, which are shown in \cref{fig:cw-all-sky}. Each curve corresponds to a different assumed distance reach, while the gray-shaded regions denote nonphysical values of ellipticity and $\zeta$. Such projected constraints indicate that, again, higher \gwh frequencies are more easily probed than lower ones, and that sources 1-10 kpc away can only be probed at the higher frequency regime, though they would require either large ellipticities ($\varepsilon>10^{-2}$) or a large fraction of \gwh power to go into dipolar \gwh radiation. Sources under 0.1 kpc from us can be probed with even tinier ellipticities and $\zeta$, as small as $10^{-5}$ in both cases. For comparison, only half of the 22 \puls probed in \citep{LIGOScientific:2021hvc} can constrain $\zeta<1$, and of those, only three can probe $\zeta<10^{-3}$ at $d\sim 0.1$ kpc. This implies that searches in \D will be able to heavily constrain both highly deformed \coss and the \BD modified theory of gravity.

\section{Population of detectable neutron stars in all-sky searches}\label{sec:allsky}

In addition to the projected upper limits in Sec. \ref{sec:cw-lims} for targeted and all-sky searches that indicate tight constraints on ellipticity and the \BD parameter, it is also worthwhile to fold some astrophysics into understanding how many sources we would expect to detect in \D. Such a question inherently relies on assumptions regarding the spin frequency and spatial distributions of \nss, which may or may not be reflected in the \puls we observe electromagnetically.

We estimate the number of in-band sources we can detect with \D following the same approach as \citep{Reed:2021scb} in the complementary frequency range between 0.1 and 10 Hz. We briefly recall the main assumptions during the calculation. 1) We calculate the probability distribution function of the \gwh frequency $\Phi(f_{\rm gw})$ from the ATNF catalog \citep{Manchester:2004bp} \pul spin distribution using $f_{\rm gw}=2 f_{\rm NS}$. This translates into approximately 71\% of the Galactic \nss emitting \gws in the frequency band that we consider here. 2) We assume the spatial distribution of Galactic \nss to follow an exponential distribution in the vertical direction above the Galactic disk and a Gaussian-like distribution in the radial direction from the Galactic Center \citep{binney1998galactic, Faucher-Giguere:2009fex,Taani:2012xp}, and express it as a function of the\nsh distance from Earth $d$
\begin{widetext}
\begin{equation}
    \label{eq:galactic_pulsar_spatial_distribution}
    \rho(d) = \frac{N_0 d^2}{\sigma_r^2 z_0} \int_0^1 \exp\left( - \frac{x d }{z_0} \right) I_0 \left( \frac{R_e d \sqrt{1-x^2}}{\sigma_r^2}\right) \exp\left(-\frac{R_e^2 +d^2\left( 1 -x^2\right)}{2 \sigma_r^2}\right) \mathrm{d}x,
\end{equation}
\end{widetext}
where $I_{0}$ is the modified Bessel function, $N_{0}$ is the number of \nss in our Galaxy, $R_{e}$ is the distance from the Galactic Center to Earth, $\sigma_{r}$ a radius parameter, and $z_{0}$ is the thickness of the Galactic disk. Note that the integral of $\rho(d)$ from 0 to $d$, defined as $N(d)$, is normalized to $N_{0}$.

\begin{figure}[htbp]
  \centering
  \includegraphics[width=\columnwidth]{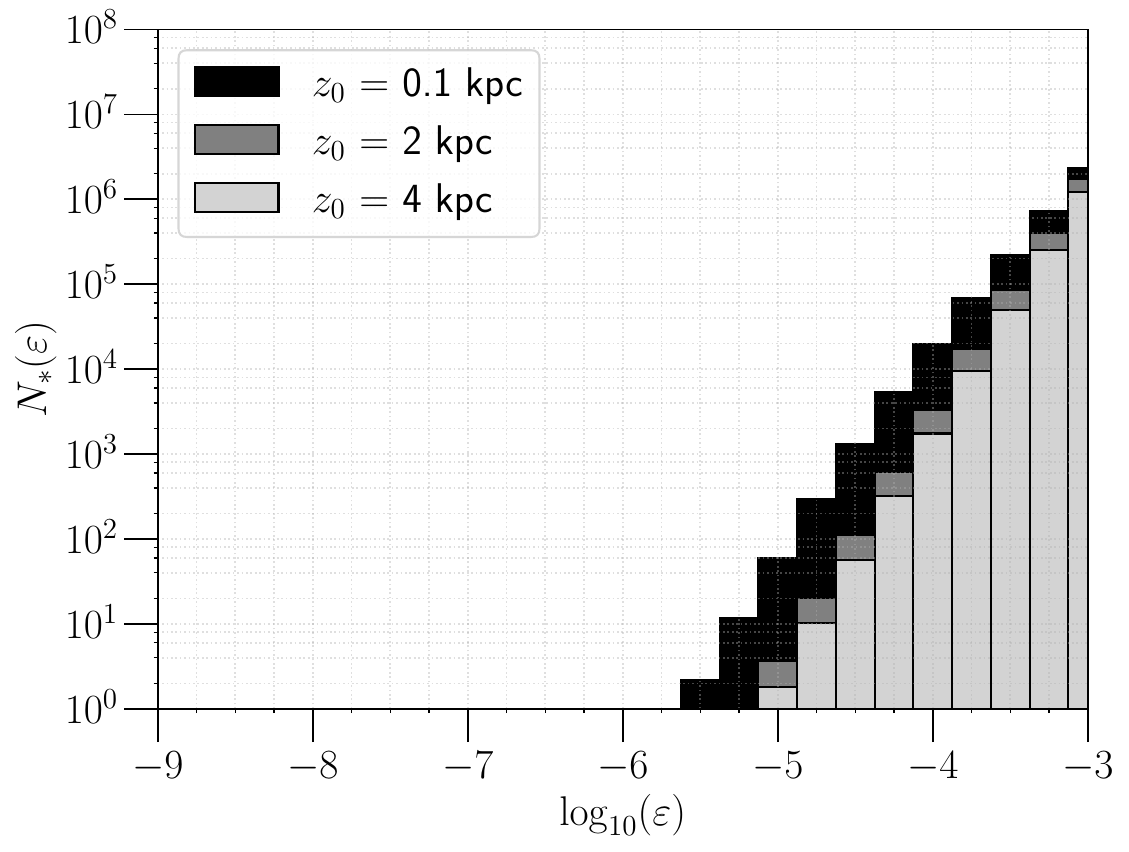}
  \caption{Number of detectable isolated \nss in an all-sky search as a function of ellipticity, assuming $\Tobs=1$ year, $\TFFT=1$ day, $N_{0} =10^{8}$, and the spatial, distance and frequency distributions of \nss in the galaxy as in the main text and integrating over them. Different $z_0$ represent different thicknesses of the galactic disk.}
  \label{fig:nstar-vs-ellip}
\end{figure}

Given the two previous assumptions, the number of detectable \ns at a given ellipticity $\varepsilon$ can be expressed as \citep{Reed:2021scb}:
%\begin{equation}
%    \label{eq:N_star_eps}
%    N_{*}(\varepsilon) = \int_{\log_{10}f_{\rm min}}^{\log_{10}f_{\rm max}} N(d(\log_{10}(f_{\rm gw}), \, \varepsilon)) \Phi(\log_{10}(f_{\rm gw}))\mathrm{d}\log_{10}(f_{\rm gw}).
%\end{equation}
\begin{equation}
    \label{eq:N_star_eps}
    N_{*}(\varepsilon) = \int_{0.1\, \mathrm{Hz}}^{10\, \mathrm{Hz}} N\left(d(f_{\rm gw}, \, \varepsilon)\right) \Phi(f_{\rm gw})\mathrm{d}f_{\rm gw}.
\end{equation}
%LEAVE THIS COMMENT HERE FOR NOW!
%The above equation tacitly assumes that the \nss have all the same ellipticity, and hence it will lead to conservative \textbf{(overestimated?)} predictions. %\tcr{Think whether to add by hand some ellipticity distribution, e.g., log-uniform distribution.} 
Given three different values of disk thickness $z_{0} = \{0.1,\, 2,\, 4\}\, \mathrm{kpc}$ \citep{Reed:2021scb}, $R_{e}=8.25 \, \mathrm{kpc}$ \citep{Gravity:2019nxk}, $\sigma_{r}= 5\, \mathrm{kpc}$ \citep{Faucher-Giguere:2009fex}, and $N_{0}=10^{8}$, we illustrate the corresponding $N_{*}(\varepsilon)$ in \cref{fig:nstar-vs-ellip}. 
It appears that \D will be able to prove the existence of a very high number of \nss ($\sim [10^{4}-10^{6}$]) only if their ellipticity $\varepsilon \sim \mathcal{O}(10^{-4}-10^{-3})$. Interestingly, these high degrees of deviation from spherical symmetry are not accessible to ground-based GW detectors with current \cwh searches due to their limitations on $\dot{f}_{\rm gw}$. This shows once more the complementary of the \cwh physics from \D in the $[0.1,\,  10]$ Hz band with respect to the ground-based GW detectors' one in the $[20,\, 2000]$ Hz band.

\section{Probing highly eccentric double neutron-star systems}\label{sec:nsbin}

\begin{figure*}[ht!]
     \begin{center}
        \subfigure[ ]{%
            \label{fig:binary-region-probable}

        \includegraphics[width=0.5\textwidth]{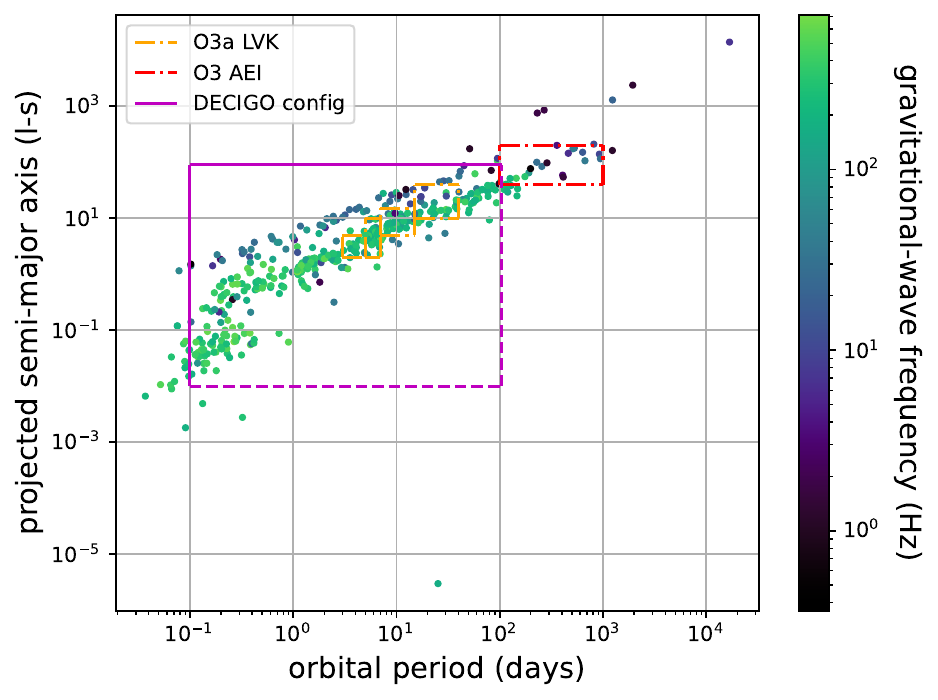}
        }%
        \subfigure[]{%
           \label{fig:binary-max-ecc-probable}
           \includegraphics[width=0.5\textwidth]{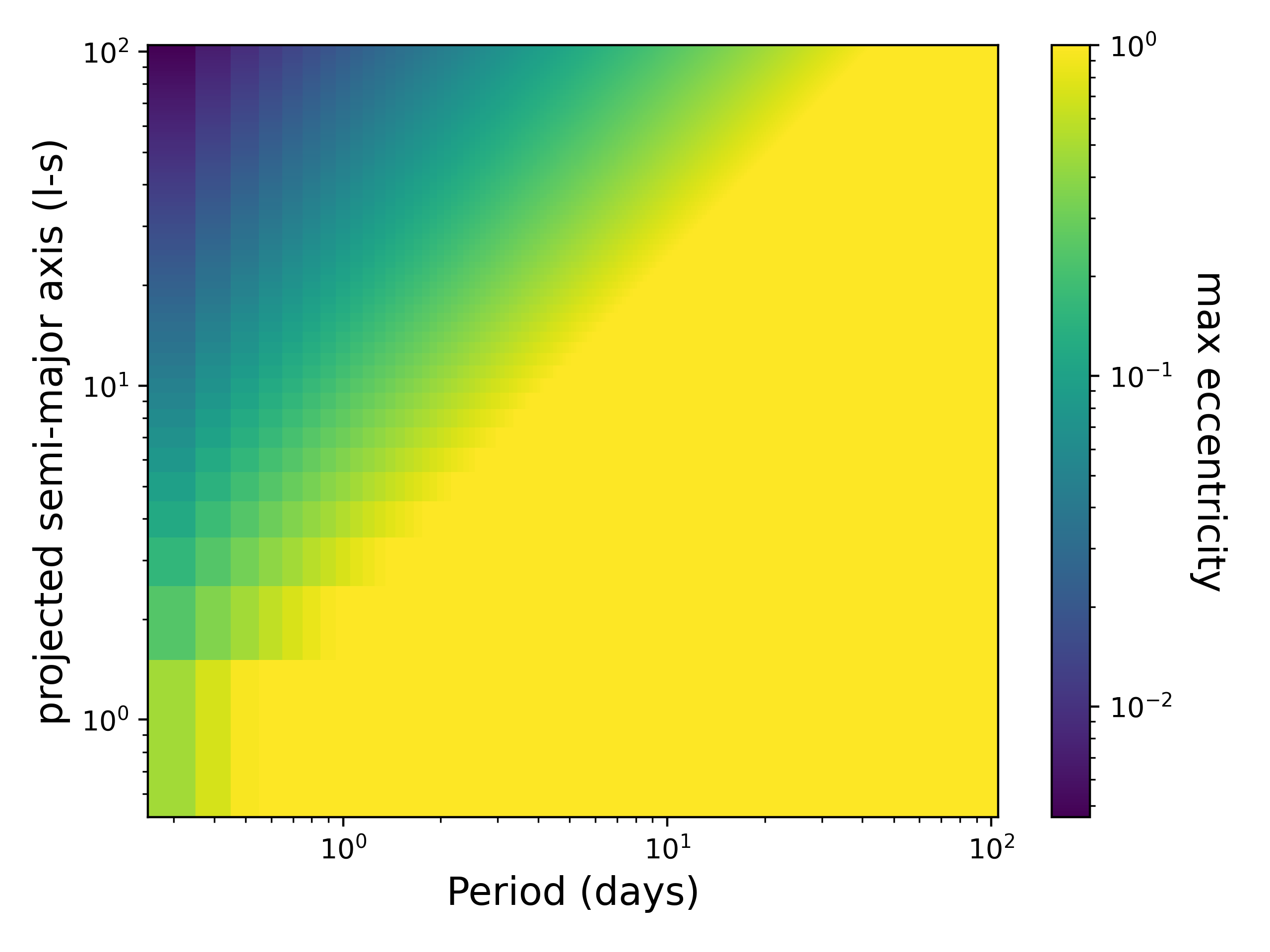}
        }\\ %  ------- End of the first row ----------------------%
    \end{center}
    \caption{(a) Binary parameter space that can be probed in \D relative to those in current ground-based \gwh detectors. The purple region represents one realization of the orbital parameter space that can be searched over in \D that has an equivalent computational cost as one of the orange configurations. Known \puls are plotted as dots, though we note that most \puls here have been detected at frequencies above the \D frequency band. The dashed purple lines indicate that the region can be extended to lower $a_p$ or higher periods with negligible additional computational cost. (b) Maximum possible eccentricity of a binary that could be detected in a hypothetical all-sky search for binary systems without explicitly searching over the eccentricity parameter. We consider $\fgw=1$ Hz to make this plot in \cref{eq:eccmax}.} 
     \label{fig:binary-all-sky-search}
\end{figure*}

\Nss in binary systems are promising sources of \cws, but in ground-based detectors, searching for them requires an excessive amount of computing power, which limits the chosen $\TFFT$ and orbital parameter space. In \D, however, the low-frequency range would allow us to perform all-sky searches covering the same sky positions and orbital parameters in a fraction of the time, i.e. costing a few core-hours instead of $10^7$ core-hours (see App. \ref{app:compcost}). Put in another way, if we employ the same amount of computing power in \D searches as we do in current searches, we could probe a much wider orbital parameter space. \cref{fig:binary-region-probable} shows a comparison of what semi-major axis / orbital period ranges we could analyze in \D compared to those in existing all-sky searches for binary systems. While the \gwh frequencies of \nss in known binary systems from the ATNF catalog tend to lie outside the \D band, it is still worthwhile to explore unknown sources at low frequencies whose orbital parameters lie well outside those currently searched for. Furthermore, we see that larger semi-major axes and smaller orbital periods can be probed more easily with \D, since the orbital Doppler shift scales with the \gwh frequency. 

Analyzing \gws at deci-hertz frequencies would also permit probing highly eccentric systems without the need to explicitly search over this parameter. Ensuring that the signal frequency modulation is confined to one frequency due to an eccentric binary for the entire observation run, the following equation for the maximum eccentricity to which we would be sensitive is \citep{Covas:2024nzs}:

\begin{equation}
  e_{\mathrm{max}} = 0.48\,\left({1\,\textrm{Hz}\over{f_0}}\right) \left({{P_\mathrm{orb}} \over 10\,\textrm{days}}\right) \left({50\,\textrm{l-s}\over{a_p}}\right).
  \label{eq:eccmax}
\end{equation}
In \cref{fig:binary-max-ecc-probable}, we show the maximum eccentricity that we could be sensitive to as a function of the orbital parameter space. For most of this space, even down to small orbital periods, we need not worry about eccentricity when performing an all-sky search for \nss in binary systems. This contrasts greatly with current \gwh searches, that can only probe eccentricities less than $\sim 0.1$ \citep{Covas:2024nzs}.

% \subsection{Decreased computational cost for low-frequency searches for isolated NS}

\section{Projected sensitivities from stochastic \gwh searches}\label{sec:sgwb}
In light of the large number ($N_{0}\sim 10^{8}-10^{9}$) \citep{sartore2010galactic} of \nss in our Galaxy, the superposition of the persistent and weak \cwh signals from individually-undetectable ones is likely to give rise to a stochastic gravitational-wave background (SGWB) \citep{Allen:1997ad,Romano_Cornish_review_Romano:2016dpx,SGWB_astro_review_Regimbau:2011rp, SGWB_cosmo_review_Caprini:2018mtu}. The detection and characterization of such an SGWB would provide constraints that are independent and complementary to those inferred from \cwh (and electromagnetic) searches for individual NSs. In addition to that, the SGWB would give immediate insight into the ensemble properties of NSs, by shedding light on the statistical distributions of the parameters (e.g., the \nsh ellipticity and its mean value) characterizing the Galactic \nsh population.

SGWB searches typically characterize the fractional energy density $\Omega_{\mathrm{GW}}$~\citep{Allen:1997ad,Romano_Cornish_review_Romano:2016dpx}, which is defined as the ratio between $\rho_{\mathrm{GW}}$, the energy density from all GWs in the Universe, and $\rho_{c} \equiv \frac{3 H_0^2 c^2}{8 \pi G}$, the critical density needed to have a flat Universe. Here, $H_0 = 67.9 \, \mathrm{km\, s^{-1} \, Mpc^{-1}}$~\citep{Planck:2015fie} is Hubble's parameter today.
Given that $\Omega_{\mathrm{GW}}$ receives contributions from GWs at all frequencies, it is natural to study its frequency spectrum

%\begin{equation}
%\label{eq:S_h_spectral_shape}
%    S_{h}(f_{\rm gw}) = \frac{32 \pi^4 G^2 \left\langle \varepsilon ^2 \right\rangle_{\mathrm{NS}} \left\langle I_{zz}^2 \right\rangle_{\mathrm{NS}}}{5 c^8}\, \left\langle \frac{1}{d^2} \right\rangle_{\mathrm{NS}}\, f_{\rm gw}^4 \, N_{0}\Phi(f_{\rm gw}) \,,
%\end{equation}

\begin{equation}
    \label{eq:omega_f_definition}
    \Omega_{\mathrm{gw}}(f_{\rm gw}) = \frac{f_{\rm gw}}{\rho_{c}} \, \frac{d\rho_{\mathrm{gw}}(f_{\rm gw})}{df_{\rm gw}}.
\end{equation}
For an ensemble of \puls, whose contributions are summed incoherently, the expression for $\Omega_{\mathrm{gw}}(f_{\rm gw})$ is \citep{SGWB_NS_ellipticity_DeLillo:2022blw}
\begin{align}
    \label{eq:omega_f_pulsars}
    \Omega_{\rm gw}(f_{\rm gw}) &= 
    \frac{64 \pi^6 G^2}{3 H_0^2} \frac{\left\langle \varepsilon ^2 \right\rangle_{\mathrm{NS}} \left\langle I_{zz}^2 \right\rangle_{\mathrm{NS}}}{5 c^8}\,  \\ &\times \left\langle \frac{1}{d^2} \right\rangle_{\mathrm{NS}}\, f_{\rm gw}^7 \, N_{\rm band}\, \Phi(f_{\rm gw}) \,,
\end{align}
where the angular brackets $\left\langle ... \right\rangle_{\mathrm{NS}}$ denote the ensemble average over the \nsh population and $N_{\rm band}$ the number of in-band NSs.

Following the methods presented in \citep{SGWB_NS_ellipticity_Talukder:2014eba, SGWB_NS_ellipticity_DeLillo:2022blw, Agarwal:2022lvk}, we perform a sensitivity study about the \D capability to characterize an SGWB from isolated, rotating, non-axisymmetric NSs. Considering a single constellation of detectors in the triangular configuration that makes use of time-delay interferometry, it is not possible to apply the cross-correlation techniques that one uses with ground-based GW interferometric detectors, and hence one must rely on autocorrelation-based methods \citep{Tinto:2001ui, Hogan:2001jn} and Bayesian model selection \citep{Baghi:2023qnq} to search for an SGWB. The following projections assume such a setup, and the results refer to the case where a perfect subtraction of instrumental noise and/or any unwanted astrophysical foregrounds, mainly from \nsh binaries (but also black hole binaries and black hole \nsh binaries) \citep{Cutler:2005qq, Harms:2008xv}, happens.

\begin{figure*}[ht!]
     \begin{center}
        \subfigure[ ]{%
            \label{fig:PI-curve}

        \includegraphics[width=0.5\textwidth]{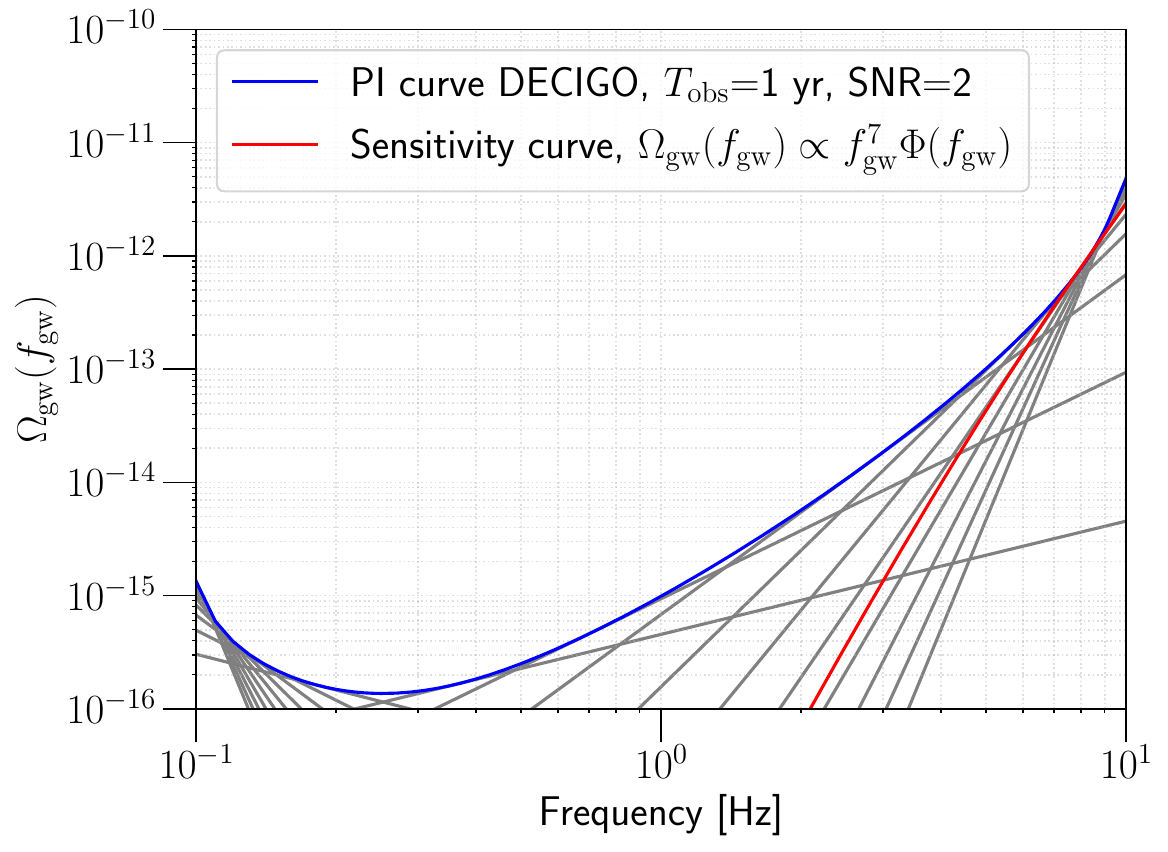}
        }%
        \subfigure[]{%
           \label{fig:avg-ellip-vs-N0}
           \includegraphics[width=0.5\textwidth]{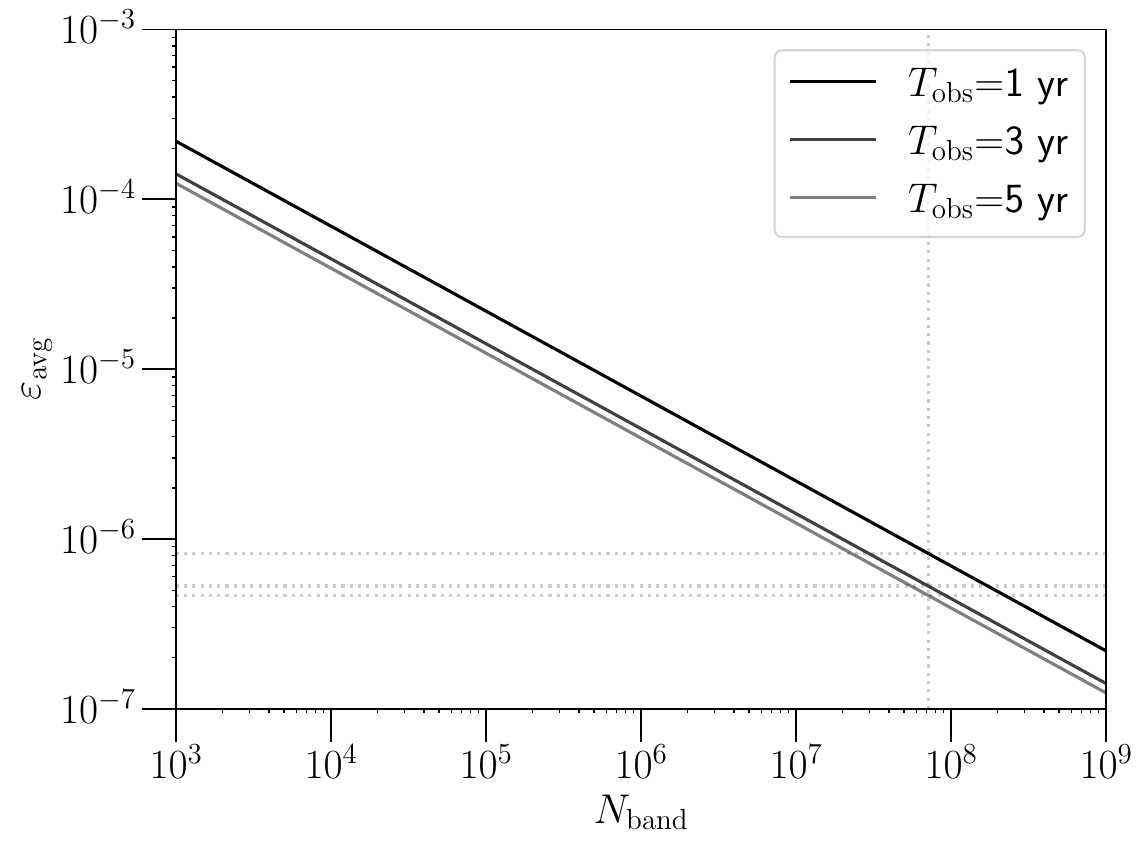}
        }\\ %  ------- End of the first row ----------------------%
    \end{center}
    \caption{(a) 2-sigma PI curve for \D (blue line, solid), assuming one year of observation in the [0.1, 1]-Hz band and perfect subtraction of instrumental noise
and/or any unwanted astrophysical foreground; individual power-law sensitivity curves (gray, solid); and sensitivity curve for $\Omega_{\rm gw} \propto f^{7}\, \Phi(f)$ (red, solid). (b) Average ellipticity of the \nsh population detectable as a function of the number of \nss in the galaxy assuming an observation time of 1, 3, and 5 years (black, gray, and light-gray lines, respectively).}
     \label{fig:stoch-gwb-proj}
\end{figure*}

In \cref{fig:PI-curve}, we present the power-law integrated (PI) curve \citep{PI_curves_Thrane:2013oya} for \D, assuming the analytical \psd from \citep{Yagi:2011wg}, one-year observation time and a \snr of two. The PI curve represents the sensitivity of an SGWB search assuming that $\Omega_{\rm gw}(f)$ follows a power law in frequency and is the convolution of the sensitivity curves for each power law (gray curves in \cref{fig:PI-curve}). In spite of the non-trivial dependency of frequency distribution of the \puls $\Phi(f_{\rm gw})$, the corresponding sensitivity curve (red curve in \cref{fig:PI-curve}) is still well-approximated by a power law.

Given the sensitivity curve for a \nsh population, we specialize equation \eqref{eq:omega_f_pulsars} to the case of Galactic \nss (with $\left\langle 1/d^2\right\rangle^{-1/2}= 6\; \mathrm{kpc}$ and $\left\langle I_{zz}^2 \right\rangle_{\mathrm{NS}}^{1/2} = 10^{38}\, \mathrm{kg\,  m^2}$), and invert it to evaluate the average ellipticity
$\varepsilon_{\rm av} (N_{\rm band}) \approx \sqrt{\left\langle \varepsilon ^2 \right\rangle_{\mathrm{NS}}}$ \citep{SGWB_NS_ellipticity_Talukder:2014eba, SGWB_NS_ellipticity_DeLillo:2022blw} as a function of the number of \nss emitting in the \D frequency band. We illustrate such a curve in \cref{fig:avg-ellip-vs-N0} for observation times of 1, 3, and 5 years observation times. Taking as a reference number around $10^{8}$ in-band \nss, searches for SGWB from isolated \nss in our Galaxy with \D will be able to probe a population with average ellipticity down to $[5-9]\times 10^{-7}$ with an signal-to-noise ratio of two in one to five years observation time.

\section{Multi-detector \gwh astronomy with \D and \et}

Both \D and \et \cite{Punturo:2010zz} will be sensitive to \gwh frequencies below 10 Hz, necessitating some comparison of how well they could perform to detect both known and unknown \coss. Such projections for \cwh searches were already performed in \cite{Pitkin:2011sc}, emphasizing potential constraints on \eos and the magnetic fields of \puls. However, here we present the possibility of doing multi-band \gwh astronomy with \D and \et. The coexistence of such detectors would allow us not only to confirm (or reject) the detection of \coss via coincident observation,  but also enhance the sensitivity of an eventual detection. 

The same benefits of doing \cwh searches with \D  discussed in this paper -- affordable computational costs, sensitivity to low-frequency known and unknown \coss, and sensitivity to eccentric binaries -- would apply as well to \et, although this will depend on the eventual minimum frequency to which \et could probe, namely between [2,5] Hz.

\begin{figure}[htbp]
  \centering
  \includegraphics[width=\columnwidth]{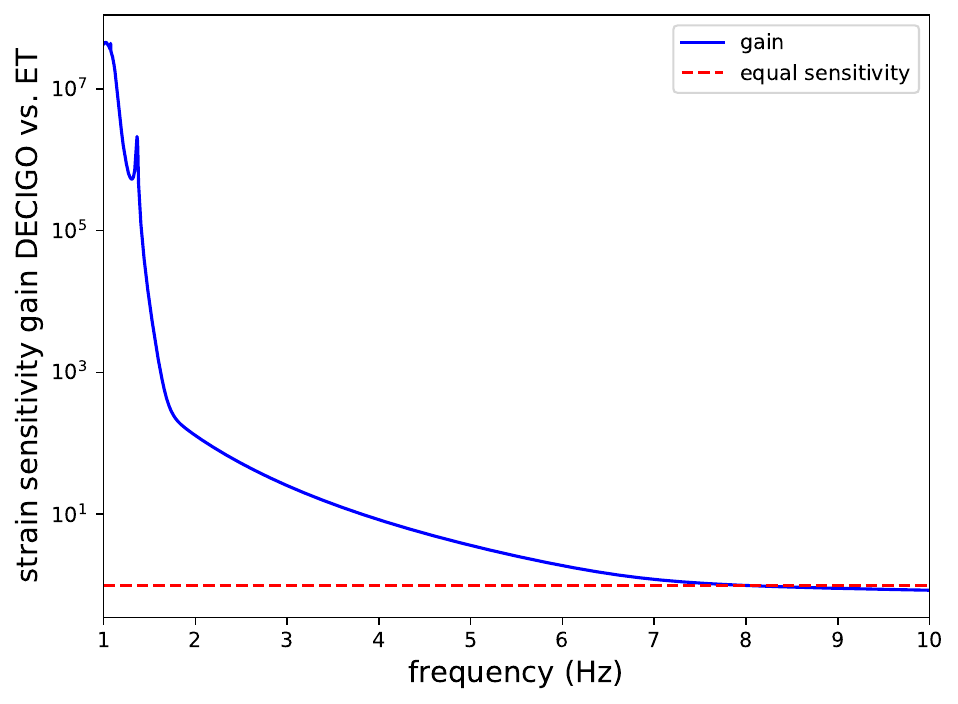}
  \caption{The gain in sensitivity in \cwh searches by using \D versus \et, computed by taking the square root of the ratio of the two \psds. The projected constraints for \D in \cref{fig:cw-targ} and \cref{fig:cw-all-sky} can be multiplied by this gain factor to pass to projected constraints in \et. }
  \label{fig:decigo-vs-et-sens}
\end{figure}

In \cref{fig:decigo-vs-et-sens}, we compute the nominal sensitivity gain of \D with respect to \et for searches for \cws from \coss. Over most of the low-frequency range, \D will be much more sensitive to these kinds of \coss compared to \et. Additionally, \D will probe \coss below 2 Hz, which comprise most of the ATNF catalog sources, something that \et likely cannot do, and even if it could, its sensitivity degrades with respect to that of \D at such low frequencies. The sensitivity gain $\mathcal{G}$ is simply the ratio of the \psds of \D and \et: $\mathcal{G}\equiv\sqrt{S_{\rm DECIGO}(f) / S_{\rm ET}(f)}$, and is a factor of at least 10 at all frequencies below $\fgw\lesssim 5$ Hz, and approaches no gain ($\mathcal{G}\rightarrow 1$) at $\fgw\simeq 8$ Hz. The strain sensitivity is directly proportional to the ellipticity, so \D will probe ellipticities about a factor of 10 smaller than those in \et at frequencies below 5 Hz.

\cref{fig:decigo-vs-et-sens} quantifies the expected gain in strain sensitivity using \D compared to \et for \cwh searches, and the results in \cref{fig:cw-targ} and \cref{fig:cw-all-sky} can be simply decreased by $\mathcal{G}$ to interpret them as arising from future \et data.

\begin{figure}
    \centering
    \includegraphics[width=\columnwidth]{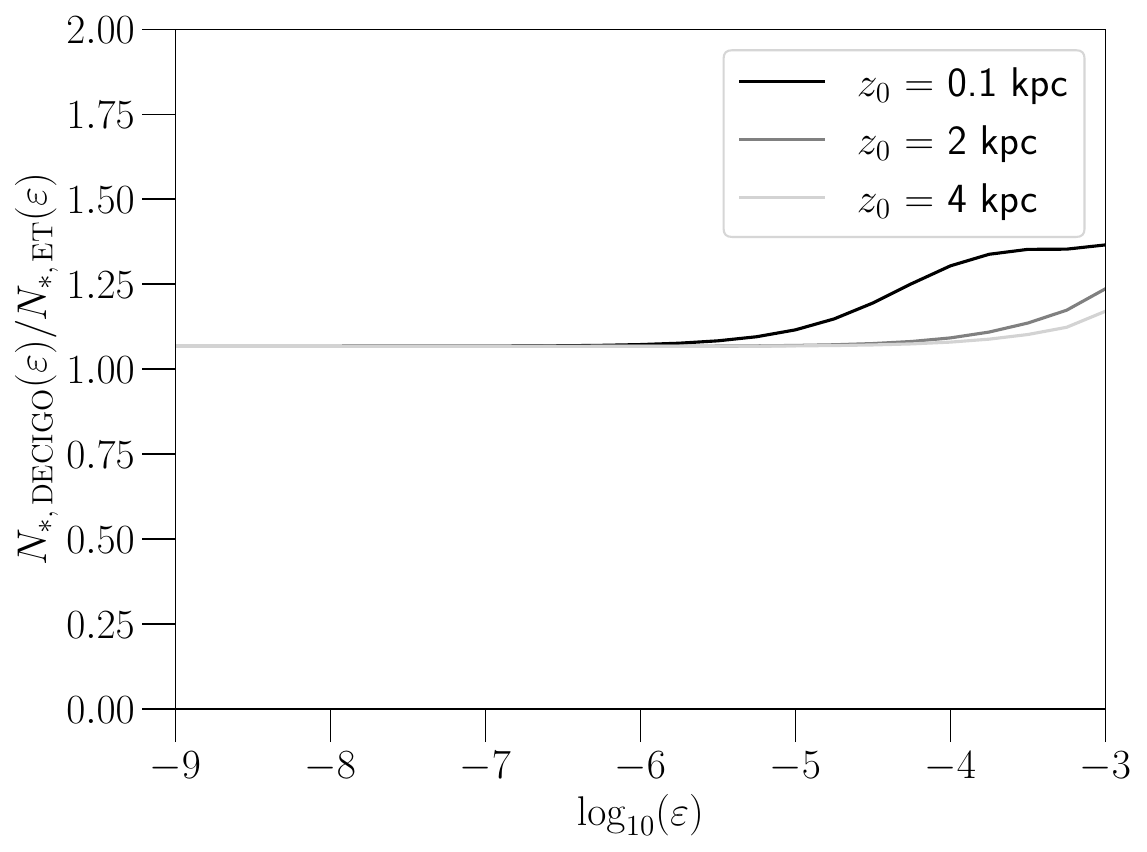}
    \caption{Ratio of the number of observable \coss with \D to the \et one as a function of the elliptictity in the same framework as \cite{Reed:2021scb} and for different thicknesses of the galactic disk $z_{0}$.}
    \label{fig:DECIGO_vs_ET_Nstar}
\end{figure}

The difference in the \psd has less straightforward implications when one performs population studies in the style of \citep{Reed:2021scb} and \sgwb studies, given the broad-band nature of these studies in contrast to the narrow-band \cwh searches. As a consequence, one must be careful to make a fair comparison, since having a broader (narrower) band translates into a larger (smaller) fraction of the population that contributes to the number of detectable \coss, and in an overall increased (decreased) sensitivity, as would be seen in the the PI curve, for \sgwb searches \citep{PI_curves_Thrane:2013oya}. When we compare our results in terms of $N_{*}(\varepsilon)$, we find that \D would detect at most $\sim[5,30]\%$ more \coss than \et, as shown in \cref{fig:DECIGO_vs_ET_Nstar}. A similar result holds for \sgwb searches, where \D appears to be 1.35 times more sensitive than \et to $\varepsilon_{\rm av} (N_{\rm band})$. The trend is instead the opposite if one considers the complementary [10-2000]-Hz band, where \et is noticeably more sensitive than \D and able to constrain the \gwh signal from sources with a much lower ellipticity.

\section{Conclusions}\label{sec:concl}

% Multi-band with ET, enhancement of sensitivity using both detectors; Number of detectable pulsars versus frac of energy going assuming ellip distirubtion; $\sqrt{5}$ DECIGO PSD

In this work, we have assessed the capability of \D to characterize the GWs from highly deformed \coss, with a particular focus on \cwh searches. \tcb{Detecting \gws from such objects, whose existence has little evidence today, would offer an avenue for new, unexpected discoveries}. We find that \D would be capable of operating in a complementary way to ground-based GW detectors. First, it would allow us to explore the $[0.1, \, 10]$ Hz band, hence opening a new window on multi-band GW astronomy of \cwh searches with present and future ground-based GW detector networks. Second, \cwh searches would be free from the first-frequency-derivative constraints that limit the sensitivity of ground-based GW detectors' \cwh searches, allowing to probe extreme ellipticity regions otherwise inaccessible. Third, performing the same kinds of searches in \D that are currently done for isolated \nss and those in binary systems would require a fraction of the computational cost, permitting exhaustive coverage of orbital parameter space that is currently impossible now. Fourth, if the Galactic \puls follow the rotational frequency distribution from the ATNF catalog, \D would investigate the properties of the part of the Galactic \pul population to which ground-based detectors are blind. Fifth, even if \D \cwh searches were not able to individually detect the majority of the Galactic \coss, it would still be possible to search for the SGWB arising from the superposition of their weak and persistent \cwh signals to gain information about their ensemble properties at once. And finally, \D would probe a similar low-frequency band as in \et, although with a slightly enhanced sensitivity, which would allow the possibility of coincident detection of \cwh and \sgwb sources.

The results here forecast that \D will be able to heavily constrain both highly deformed \coss and the \BD modified theory of gravity by means of targeted and all-sky \cwh searches.  Moreover, we expect \D to give insightful information about the Galactic \co population. In fact, it will either detect hundreds of thousands of highly deformed \coss, hence proving their existence, or constrain their ensemble properties by measuring the resulting SGWB, in just one year of observation.

The above motivations and results make it clear how much \D would be important to unveil and study GW emissions from Galactic \coss.

\section*{Acknowledgments}
This material is based upon work supported by NSF's LIGO Laboratory which is a major facility fully funded by the National Science Foundation

We would like to thank Gopalkrishna Prabhu and David Keitel for their comments on our draft.

Data and codes to create the plots for this analysis are available \cite{miller2025decigocw}.

\appendix

\section{Computational costs}\label{app:compcost}

We show here how computational costs for a \D search for isolated \nss and those in binaries scale with frequency and orbital parameters. To obtain a sense of how computationally efficient searches in \D will be, we determine the possible parameter space probed with the computational cost of all-sky \cwh searches in O3 of $\sim 10^7$ core-hours \cite{KAGRA:2022dwb}, and also show how the same searches currently performed on \lvk data could be done in only a few hours with \D data.

\begin{figure*}[ht!]
     \begin{center}
        \subfigure[ ]{%
            \label{fig:core-hours-isolated-all-sky}

        \includegraphics[width=0.5\textwidth]{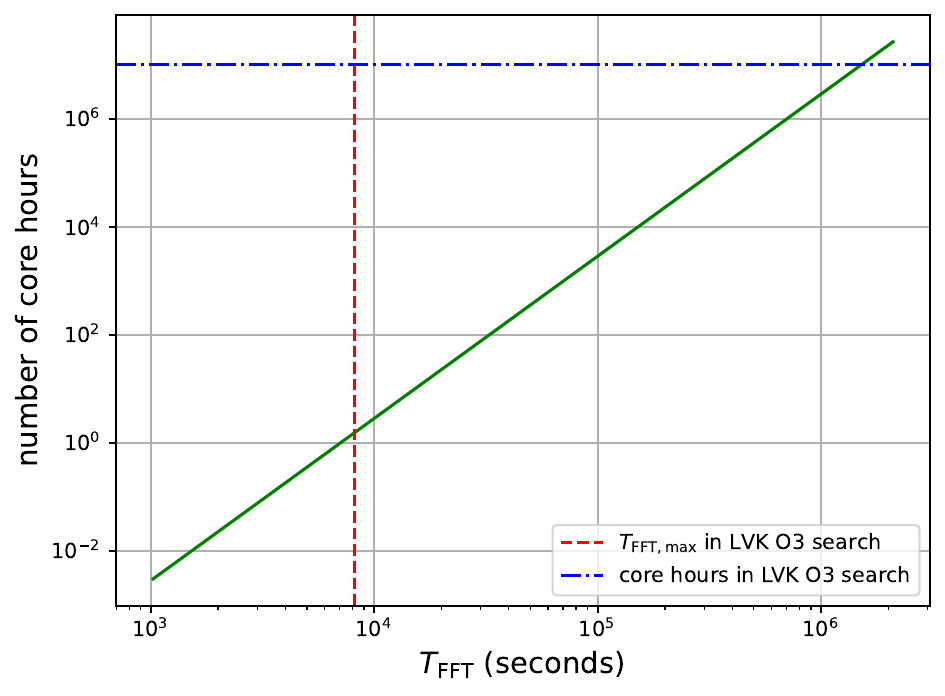}
        }%
        \subfigure[]{%
           \label{fig:num-temps-binary-all-sky}
           \includegraphics[width=0.5\textwidth]{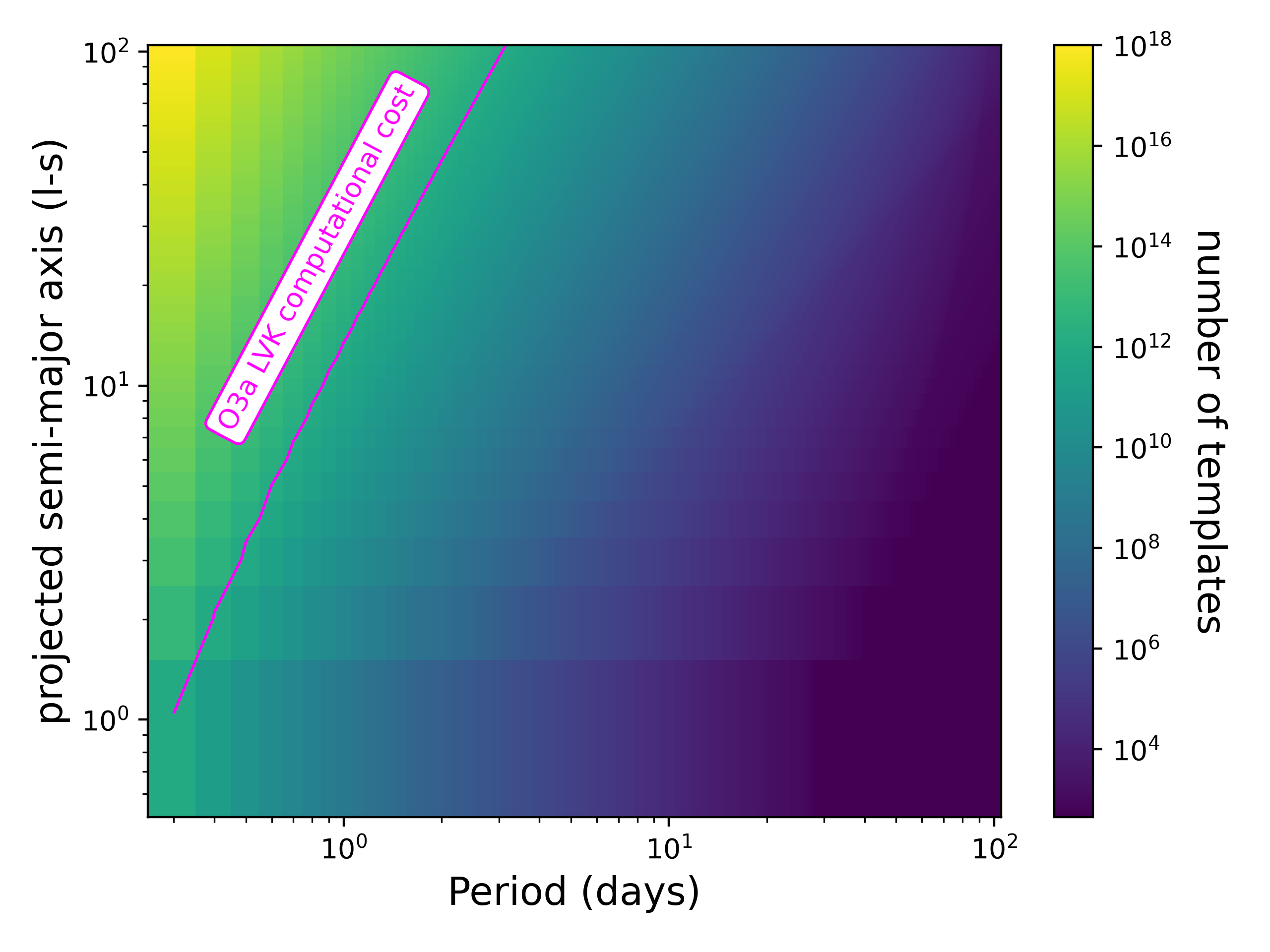}
        }\\ %  ------- End of the first row ----------------------%
    \end{center}
    \caption{(a) Computational cost in core-hours of an all-sky search for isolated \ns in \D as a function of $\TFFT$. This cost is determined by computing the number of sky points needed in the 0.1-10 Hz band at the different $\TFFT$, and noting that in O3, the computational cost of analyzing the 10-2048 Hz band was $10^7$ core hours for $10^{11}$ points in the frequency, spin-down and sky position parameter space, and assuming that each point requires the same amount of computing power to search over. (b) Number of templates in a search for \nss in binary systems as a function of the ranges of orbital period and semi-major axis of the binary. The period ranges from the value on the $x$-axis to 105 days, while the semi-major axis ranges from $10^{-2}$ l-s to the value on the $y$-axis. This plot does not include the number of sky points to search over, though we note for the chosen $\TFFT=1024$ s, as shown in the left-hand figure, the computational cost to search over the whole sky is negligible.} 
     \label{fig:comp-costs}
\end{figure*}

At the computational cost used for the searches in \cite{KAGRA:2022dwb}, ($10^7$ core hours), we can use a $\TFFT$ of 1.5 days, allowing for a nominal sensitivity improvement of $(86400*1.5/8192)^{1/4} = 4$ in strain amplitude, as shown in \cref{fig:core-hours-isolated-all-sky}. Unfortunately, this sensitivity gain is balanced by four orders of magnitude in degradation of the \gwh amplitude ($\propto f^2)$ when moving from, say 100 Hz to 1 Hz. That is why \cwh searches would be sensitive to higher-ellipticity sources.  

Likewise, if we utilize the same $\TFFT$ as used in the searches in \cite{KAGRA:2022dwb} (red-dashed line in \cref{fig:core-hours-isolated-all-sky}), we can see that the search would be performed in under one hour. The computational cost of performing all-sky searches for \cws is therefore significantly reduced, primarily due to the fact the number of points in the sky scales with $\fgw^2$ \cite{Prix:2009oha}. 

Current all-sky searches use grids on $\fdotgw$ to allow for the possibility of a small but measurable spin-down of \nss. However, in \D, the grid on $\fdotgw$ has negligible cost, since, as shown in \cref{eqn:fdotgw}, for even large ellipticities, the spin-down is small enough to be contained within one frequency bin. Thus, this is an additional computational gain with respect to current all-sky searches, and also one in sensitivity, since the total number of templates analyzed, and thus the trials factor, is smaller without the $\fdotgw$ contribution. 

The computational cost improvement for all-sky searches for \nss in binary systems is also significant. We show in \cref{fig:num-temps-binary-all-sky} the number of templates required to search over the semi-major axis and orbital period parameter space in \D. We draw a magenta line to indicate the number of templates used in the O3a all-sky search for binaries \cite{LIGOScientific:2020qhb}, which covered four discrete squares in this parameter space, indicated in \cref{fig:binary-region-probable}. We find that a wide range of semi-major axes and orbital periods become accessible in \D, which comprise a large number of known \pul parameters.

\bibliographystyle{apsrev4-1}
\bibliography{biblio,biblio_pbh_method}

\end{document}